\documentclass[10,fleqn,]{article}
\usepackage{amsfonts,graphicx}

\def\noi{\noindent}

\makeatletter
\renewcommand{\section}{\@startsection{section}{1}{0pt}%
        {-0.5ex plus -.5ex minus -.2ex}{0.3ex plus .2ex}%
        {\large\bf\protect\raggedright}}

\renewcommand{\subsection}{\@startsection{subsection}{2}{0pt}%
        {-3ex plus -1ex minus -.2ex}{1.4ex plus .2ex}%
        {\normalsize\bf\protect\raggedright}}

\renewcommand{\thesubsubsection}%
        {\arabic{section}.\arabic{subsection}.\arabic{subsubsection}.}
\newcommand{\para}{\@startsection{paragraph}{4}{0pt}%
        {1.5ex plus -.5ex minus -.2ex}{-1em}{\normalsize\bf}}

\renewcommand{\@oddhead}{\raisebox{0pt}[\headheight][0pt]{%
   \vbox{\hbox to\textwidth{\rightmark \hfil \rm \thepage \strut}\hrule}}}
\renewcommand{\@evenhead}{\raisebox{0pt}[\headheight][0pt]{%
   \vbox{\hbox to\textwidth{\thepage \hfil \leftmark \strut}\hrule}}}

\makeatother

\newcommand{\Title}[1]{\noi {\Large #1} \\}

\newcommand{\Abstract}[1]{\vskip 2mm \begin{center}
        \parbox{16.4cm}{\small\noi #1} \end{center}\medskip}

\newcommand{\email}[2]{\footnotetext[#1]{e-mail: #2}
    \addtocounter{footnote}{1}}

\def\para{\paragraph}

\def\beq{\begin{equation}}
\def\eeq{\end{equation}}
\def\bear{\begin{eqnarray}}

\def\ear{\end{eqnarray}}

\def\Jl#1#2{{\it #1\/} {\bf #2},\ }

\def\CQG#1 {\Jl{Class. Qu. Grav.}{#1}}
\def\DAN#1 {\Jl{Dokl. AN SSSR}{#1}}
\def\GC#1 {\Jl{Grav. \& Cosmol.}{#1}}
\def\GRG#1 {\Jl{Gen. Rel. Grav.}{#1}}
\def\JETF#1 {\Jl{Zh. Eksp. Teor. Fiz.}{#1}}
\def\JMP#1 {\Jl{J. Math. Phys.}{#1}}
\def\NPB#1 {\Jl{Nucl. Phys.}{B\ #1}}
\def\PLA#1 {\Jl{Phys. Lett.}{#1A}}
\def\PLB#1 {\Jl{Phys. Lett.}{#1B}}
\def\PRD#1 {\Jl{Phys. Rev.}{D\ #1}}
\def\PRL#1 {\Jl{Phys. Rev. Lett.}{#1}}

\begin{document}
\twocolumn[ \thispagestyle{empty}
\bigskip

\Title {\uppercase{Quantum birth of a hot universe II.\\
Model parameter estimates\\ from CMB temperature fluctuations}}

\noi{\large\bf Irina Dymnikova $^{a,1}$ and Michael
Fil'chenkov$^{a,2}$}

\medskip{\protect
\begin{description}\itemsep -1pt
\item[$^a$]{\it Department of Mathematics \\and Computer Science,
University of Warmia and Mazury,\\
\.Zo{\l}nierska 14, 10-561 Olsztyn, Poland}

\item[$^b$]{\it Friedmann
Laboratory\\ for Theoretical Physics,
St. Petersburg, Russia\\
Institute of Gravitation and Cosmology, Peoples' Friendship
University
of Russia,\\
6 Miklukho-Maklaya Street, Moscow 117198}
\end{description}}

\Abstract {We consider the quantum birth of a hot FRW universe
from a vacuum-dominated quantum fluctuation with admixture of
radiation and strings, which corresponds to quantum tunnelling
from a discrete energy level with a non-zero temperature. The
presence of strings with the equation of state $p=-\varepsilon/3$
mimics a positive curvature term which makes it possible, in the
case of a negative deficit angle, the quantum birth of an open and
flat universe. In the pre-de-Sitter domain radiation energy levels
are quantized. We calculate the temperature spectrum  and estimate
the range of the model parameters restricting temperature
fluctuations by the observational constraint on the CMB
anisotropy. For the GUT scale of initial de Sitter vacuum the
lower limit on the temperature at the start of classical evolution
is close to the values as predicted by reheating theories, while
the upper limit is far from the threshold for a monopole rest
mass.
}

]  
\email 1 {irina@matman.uwm.edu.pl} \email 2 {fil@crosna.net}

\section{Introduction}
Quantum cosmology treats quantum-mechanically the universe as a
whole and describes it by a wave function $\psi$ (for review see
\cite{vilenkin}). The full formalism of quantum geometrodynamics
was introduced in 1967 by DeWitt and applied  to a dust-filled
closed Friedmann-Robertson-Walker (FRW) universe with a
curvature-generated potential to find a discrete system of energy
levels \cite{dewitt}. In 1969 Misner extended this approach to
anisotropic cosmological models \cite{misner}.

DeWitt has calculated the energy levels in the case of  zero
cosmological term and with the boundary condition $\psi(0)=0$,
which corresponds to quantization in the well with infinite walls.
In 1972 Kalinin and Melnikov  considered the FRW closed model with
a non-zero cosmological term $\Lambda g_{\mu\nu}$, and found that
adding $\Lambda g_{\mu\nu}$ results in transformation of an
infinite well into a finite barrier \cite{KM}.

A year later Fomin \cite{F} and Tryon \cite{T}  put forward the
idea that a closed universe can be born as a quantum object from
nothing due to the uncertainty principle. In 1975 a nonsingular
model was proposed for a FRW universe arising from a quantum
fluctuation in de the Sitter vacuum \cite{us75}.
 A more detailed consideration of the origin of a universe
 in the quantum tunnelling event has been done
 in the late 70-s and early 80-s \cite{eng,atkatz,vil0,gri}.
The possibility of a multiple birth of causally disconnected
universes from the de Sitter background noticed in \cite{us75},
was investigated by Gott III for the case of an open FRW universe
\cite{gott}.

In the framework of the standard scenario, the quantum birth of
the universe is followed by decay of the de Sitter vacuum
ultimately resulting in a hot expanding universe
\cite{us75,DZS,L}. The hot model has been proved by the discovery
of the Cosmic Microwave Background (CMB) \cite{P}, first predicted
by Gamow \cite{GPR} who was also the author of the tunnel effect
in quantum mechanics \cite{GZP} basic for the quantum tunnelling
of a universe.

The wave function of the universe satisfies the Wheeler-DeWitt
equation \cite{dewitt,wheel}
$$\hat H\psi=0                                        \eqno(1)
$$
analogous to the Schr\"odinger equation. To put a universe into
the quantum mechanical context, one has to specify boundary
conditions for the wave function $\psi$. In quantum mechanics the
boundary conditions are related to the exterior of an isolated
quantum system. In case of a universe there is no exterior, and
the boundary conditions must be formulated as an independent
physical law \cite{VG}. This question has been debated in the
literature for about 15 years. The recent summary of these debates
can be found in \cite{VG}. At present there exist three approaches
to imposing boundary conditions on the wave function of the
universe: the Hartle-Hawking wave function \cite{hh}, the Vilenkin
(tunnelling) wave function \cite{V}, and the Linde wave function
\cite{lwf}.

The birth of a closed world from nothing (favoured by that its
total energy is zero \cite{F,T}) starts from arising of a quantum
fluctuation, and the probability of tunnelling describes its
quantum growth on the way to the classically permitted region
beyond the barrier confined by the values of the scale factor
$a=0$ and $a=a_0$, which implies that the eigenvalue of the
Wheeler-DeWitt operator is fixed at the energy value $E=0$
\cite{V}.

In this paper we address the question of the quantum birth of a
universe with a non-zero temperature. We apply the approach
proposed by Vilenkin for the  quantum birth of a universe from
nothing \cite{V} to the case of the quantum birth from a state
with non-zero quantized energy.

In the presence of radiation in an initial fluctuation, its energy
density in the quantized Friedmann equation written in terms of
conformal time, plays the role of an energy $E$ in the
Wheeler-DeWitt equation \cite{VG,F5,F8}.

Quantization of energy levels in the conformal time has been
investigated by Kuzmichev for the case of a closed FRW universe
filled with a scalar field and radiation and considered as a
quantum system in the curvature generated well \cite{Kz}. This
model describes the evolution of the universe as a succession of
transitions to progressively higher energy levels in the well, so
that the presently observable Universe is considered as the
quantum system in a highly excited state in accordance with the
basic idea suggested by Hartle and Hawking in 1983 \cite{hh}.

In the present paper we consider the quantum birth of a universe
from a vacuum-dominated quantum fluctuation with an admixture of
radiation and strings or some other quintessence with the equation
of state $p=-\varepsilon/3$. This corresponds to quantum birth of
a closed, flat or open universe by tunnelling from a discrete
energy level with a non-zero temperature.

In the literature the quantum birth of an open and flat universes
has been typically considered in the context of anti-de-Sitter
space-time \cite{ht,no}. In our model the nonzero probability  of
quantum birth in this case is related to the presence of strings
with a negative deficit angle which mimics the curvature term in
producing a potential appropriate for quantum tunnelling
\cite{us2000,us2001}.

\section{Model}
The FRW quantum universe is described
 by the minisuperspace model with a single degree of freedom, and
 the Wheeler-DeWitt equation reads\quad\cite{dewitt,vil0}
$$
\frac{d^2\psi}{da^2}-V(a)\psi=0                          \eqno(2)
$$
where
$$
V(a)=\frac{1}{l_{pl}^4}\left(ka^2-\frac{8\pi G\varepsilon
a^4}{3c^4}\right),
                                                                     \eqno(3)
$$
$a$ is the scale factor, $k=0,\pm 1$ is the curvature parameter.\\
In the Friedmann equations the total energy density may be written
in the form \cite{F5}
$$
\varepsilon=\varepsilon_{vac}\sum_{q=0}^6
B_q\left(\frac{a_0}{a}\right)^q.                       \eqno(4)
$$
The coefficients $B_q$ refer to contributions of different kinds
of matter. Here we chose normalizing scale $a_0$ as the de Sitter
horizon radius connected with the vacuum energy density
$\varepsilon_{vac}$ by
$$
{a_0^2}=\frac{3c^4}{8\pi G\varepsilon_{vac}}.
                                                             \eqno(5)
$$
which leads to $B_0=1$. The parameter $q$ is connected by
$$
q=3(1+\alpha)
\eqno(6)
$$
with the parameter $\alpha$ in the equation of state
$$
p=\alpha\varepsilon.
\eqno(7)
$$
For the most frequently used equations of state the parameter $q$
takes the values \cite{F5}:\\
$q=0\quad (\alpha=-1)$ for the de Sitter vacuum,\\
$q=1\quad (\alpha=-\frac{2}{3})$ for domain walls,\\
$q=2\quad (\alpha=-\frac{1}{3})$ for strings,\\
$q=3\quad (\alpha=0)$ for dust,\\
$q=4\quad (\alpha=\frac{1}{3})$ for radiation or ultrarelativistic gas,\\
$q=5\quad (\alpha=\frac{2}{3})$ for perfect gas,\\
$q=6\quad (\alpha=1)$ for ultrastiff matter.\\
Matter with a negative pressure has been recently included into
quintessence which is a time-varying spatially inhomogeneous
component of the matter content satisfying the equation of state
$p=-\alpha\varepsilon$ with $0<\alpha<1$ \cite{CDS}.

Separating a scale-factor-free term in the potential (3), we
reduce the Wheeler-DeWitt equation to the Schr\"odinger form
$$
-\frac{\hbar^2}{2m_{pl}}\frac{d^2\psi}{da^2}+(U(a)-E)\psi=0
                                                                       \eqno(8)
$$
with the energy $E$ given by
$$
E=\frac{B_4}{2}\left(\frac{a_0}{l_{pl}}\right)^2E_{Pl}
                                                                       \eqno(9)
$$
and related to the contribution of radiation to the total energy
density. Equation (8) describes a quantum system with the energy
$E$ related to radiation, in the potential created by other
components of matter content.

We consider an initial vacuum-dominated quantum fluctuation with
an admixture of radiation and strings (or some other quintessence
with the equation of state $p=-\varepsilon/3$). In this case the
potential takes the form
$$
U(a)=\frac{E_{pl}}{2l_{pl}^2}\biggl((k-B_2)a^2-\frac{a^4}{a_0^2}\biggr).
                                                                            \eqno(10)
$$
Imposing the boundary condition on the wave function at $a=0$, we
follow DeWitt who adopted $\psi(0)=0$ for a quantized FRW universe
\cite{dewitt}. At infinity we adopt the Vilenkin boundary
condition which prescribes  the presence only of the outgoing mode
of a wave function \cite{V}.

The quantization of energy in the well (a Lorentzian domain of the
pre-de-Sitter universe) is given in the WKB aproximation by the
Bohr-Sommerfeld formula\quad\cite{LQM}
$$
2\int\limits_0^{a_1}\sqrt{2m_{pl}(E_n-U)}\,da=\pi\hbar
\left(n+\frac{1}{2} \right),
                                                                                   \eqno(11)
$$
where $a_1$ is defined by $U(a_1)=E_n$. The potential (10) is
shown in Fig.1. It has  a maximum
$$
U_m=\frac{(k-B_2)^2}{8}\biggl(\frac{a_0}{l_{Pl}}\biggr)^2E_{Pl}
                                                                                   \eqno(12)
$$
at
$$
a_m=a_0\sqrt{\frac{k-B_2}{2}}
                                                                        \eqno(13)
$$
and zeros at
$$
a_3=a_0\sqrt{k-B_2}
                                                                       \eqno(14)
$$
and  $a=0$ where a potential has a minimum $U_{min}=0$.

Let us note here that for a vacuum energy scale
$E_{GUT}\sim{10^{15}}$ GeV, $a_0/l_{Pl}\sim {10^8}$. Later we
verify the validity of the WKB approximation more accurately
restricting the model parameter $k-B_2$ by the observable upper
limit on the value of the CMB anisotropy $\Delta T/T$.

Turning points $a_{1,2}$ where $U(a)=E_n$, are given by
$$
a_{1,2}^2=a_m^2\biggl(1\pm\sqrt{1-\frac{E_n}{U_m}} \biggr).
                                                                      \eqno(15)
$$
\begin{figure}
\includegraphics[width=3in,height=2in]{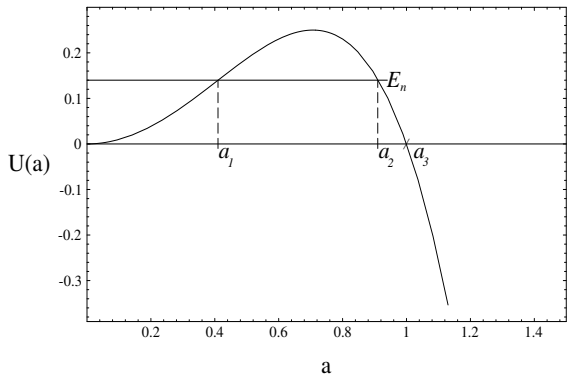}
\caption {Initial quantum fluctuation as a quantum system in the
well (a Lorentzian domain of a pre-de-Sitter universe).}
\label{fig.1}
\end{figure}
\vspace{3cm}
Approximating the potential near the maximum by
$$
U=U_m+\frac{1}{2}\frac{d^2U}{da^2}|_{a=a_m}\cdot(a -a_m)^2
                                                                    \eqno(16)
$$
and calculating the spectrum with the Bohr-Sommerfeld formula
(11), we get the general model restriction on the quantum number
$n$
$$
n+\frac{1}{2}<\frac{(k-B_2)^{3/2}}{\pi\sqrt{2}}\left
(\frac{a_0}{l_{pl}}\right)^2
                                                                     \eqno(17)
$$
By equations (4) and (9) we connect the the energy $E_n$ with the
energy density of radiation
$$
\varepsilon_{\gamma}=2\varepsilon_{vac} \frac{E_n}{E_{Pl}}
\biggl(\frac{l_{Pl}}{a_0}\biggr)^2 \biggl(\frac{a_0}{a}\biggr)^4
                                                                     \eqno(18)
$$
which is related to the temperature $\Theta=kT$ as \cite{LST}
$$
\varepsilon_{\gamma}=\frac{\pi^2}{30\hbar^3 c^3}N(\Theta)\Theta^4,
                                                                       \eqno(19)
$$
where $N(\Theta)$ counts the total number of effectively massless
degrees of freedom (species with $E\ll \Theta$) which grows with
temperature and  at the GUT scale is estimated within the range
\cite{L}
$$
N(\Theta)\sim {10^2\div 10^4}
                                                                        \eqno(20)
$$

From equation (19) we get the quantized temperature
$$
\Theta=\biggl(\frac{45}{2\pi^3N(\Theta)}\biggr)^{1/4}
\biggl(\frac{l_{Pl}}{a}\biggr)
\biggl(\frac{E_n}{E_{Pl}}\biggr)^{1/4}E_{Pl}.
                                                                      \eqno(21)
$$
The upper limit for the temperature, which follows from the basic
restriction $E_n<U_{max}$, does not depend on the value of the
parameter $k-B_2$ and is given by
$$
{\Theta}_{max}= \biggl(\frac{45}{4{\pi}^3 N(\Theta)}\biggr)^{1/4}
\biggl(\frac{l_{pl}}{a_0}\biggr)^{1/2}E_{Pl}.
                                                                         \eqno(22)
$$
For $a_0$ corresponding to the GUT scale vacuum, with $N(\Theta)$
from the range (20), the maximal possible value of the temperature
is estimated as $\Theta_{max} \approx{(0.25 \div 0.08)~ E_{GUT}}$
which is far from the monopole rest energy $E_{mon}\sim{10^{16}
\div 10^{17}}$ GeV \cite{L}.

Near the minimum the potential (10) is approximated by a harmonic
oscillator
$$
U= \frac{E_{Pl}}{2}\frac{a^2}{l_{Pl}^2}(k-B_2)
                                                                    \eqno(23)
$$
which gives the reasonable approximation up to the inflection
point $a_{infl}=a_m/\sqrt{3}$ where $U_{infl}=(5/9)U_m$.

In the region where the potential can be approximated by (23), the
energy spectrum is given by
$$
E_n=E_{pl}\sqrt{k-B_2}\left(n+\frac{1}{2}\right),
                                                                      \eqno(24)
$$
and the quantum number $n$ is restricted by the condition
$E_n<U_{infl}$ which gives
$$
n+\frac{1}{2} < \frac{5}{72}\biggl(\frac{a_0}{l_{Pl}}\biggr)^2
 (k-B_2)^{3/2}
                             \eqno(25)
$$
For this range of the quantum numbers $n$ the temperature (21)
reduces to
$$
\Theta=\biggl(\frac{45}{2\pi^3N(\Theta)}\biggr)^{1/4}
\biggl(\frac{l_{Pl}}{a_0}\biggr)
\frac{(n+\frac{1}{2})^{1/4}}{(k-B_2)^{3/8}}E_{Pl}.
                                                                       \eqno(26)
$$
We put here the value of the scale factor $a=a_2$ with which a
system starts a classical evolution beyond the barrier, to make
evident which is the dependence of the temperature on the model
parameter $k-B_2$.

 The lower limit
on the temperature corresponds to the lowest level of the energy
spectrum. This is $n=1$ for the case of the adopted boundary
condition $\psi(0)=0$, while the lowest energy possible in
principle, $E_0=\hbar \omega /2$, corresponds to $n=0$. The values
of the temperature for these two values of $n$ differs by the
factor $3^{1/4}$, and an absolute lower limit for the temperature
related to a zero-point energy $\hbar \omega /2$, is given by
$$
\Theta_{min}=\biggl(\frac{45}{4\pi^3N(\Theta)}\biggr)^{1/4}
\biggl(\frac{l_{Pl}}{a_0}\biggr) {(k-B_2)^{-3/8}}E_{Pl}
                                                                            \eqno(27)
$$

The model parameter $k-B_2$ can  be evaluated by the observational
upper bound on the CMB anisotropy \cite{S}
$$
\Delta T/T \simeq 10^{-5}.
                                                                 \eqno(28)
$$

 In the context of the inflationary
paradigm anisotropy of the relic radiation originates from  vacuum
fluctuations during inflationary stage \cite{L}. The value of
$\Delta T/T$ which arises before decay of de Sitter vacuum,
remains to be the same at the end of recombination which is
estimated in today observations \cite{lukash}.

In our model anisotropy $\Delta T/T$ at the start of a classical
evolution is related to the width of a quantized energy level of a
system inside a well. Indeed, quantum tunnelling means not a
process occurring in real time (penetration "occurs" within the
Euclidean domain where the time coordinate is imaginary), but a
nonzero probability to find a quantum system beyond the barrier
where initial quantum fluctuation starts a classical evolution, so
that no physical process affecting $\Delta T/T$ can occur "in the
course of tunnelling". As a result, the classical evolution starts
with the value of $\Delta T/T$ related to the level width in the
well which survives till the end of recombination (any additional
anisotropy appearing in the course of vacuum decay and later, is
proportional to $N_{\gamma}^{-1/2}$, where $N_{\gamma}$ is the
number of photons which can only grow in processes of decay). The
energy $E_n$  and the value of the scale factor $a_2$ (see Fig.1)
affect $\Delta T/T$ at the beginning of the classical evolution
which starts with those parameters as the initial values.

For the system at the quantum level $E_n$ the temperature
fluctuations $\Delta T$ originate from the natural width of a
level, $\Delta E_n$, and from $\Delta T$ due to statistical
 fluctuations in the photon ensemble.

Statistical fluctuations in the temperature of the
ultrarelativistic gas give \cite{LST}
$$\biggl(\frac{\Delta T}{T}\biggr)_{st}=\biggl(\frac{15}{2\pi^2
N(\Theta)}\biggr)^{1/2} \biggl(\frac{\hbar c}{\Theta}\biggr)^{3/2}
V^{-1/2}.
                                                                           \eqno(29)
$$
Putting $V=E_n/\varepsilon_{\gamma}$ and $\Theta$ from Eq.(21) we
get
$$
\biggl(\frac{\Delta T}{T}\biggr)_{st}=\frac{1}{2}
\biggl(\frac{45}{2\pi^3 N(\Theta)}\biggr)^{1/8}
\biggl(\frac{E_{Pl}}{E_n}\biggr)^{3/8}
\biggl(\frac{l_{Pl}}{a}\biggr)^{1/2}.
                                                                      \eqno(30)
$$
 The general constraint (17) and the observational constraint (28)
 restrict the model parameter $k-B_2$ by
$$
  k-B_2 > (2.6\div 4.7)\cdot 10^{-6}.
                                                             \eqno(31)
$$
Two values correspond to the range (20) for $N(\Theta)$. This
gives rough estimate by the order of magnitude, since the number
of massless degrees of freedom $N(\Theta)$ is estimated roughly up
to two orders of magnitude \cite{L}.

 The natural width $\Gamma_n=\Delta E_n$ can be
 evaluated by the level width $\Gamma_n$ for a harmonic
oscillator \cite{LQM}
$$
\Delta E_n=\Gamma_n=\frac{2\alpha}{3}\frac{\hbar\omega}{E_{Pl}}
\hbar\omega n.
                                                                         \eqno(32)
$$
From eq.(21) we get
$$
\frac{\Delta T}{T}=\frac{1}{4}\frac{\Delta E_n}{E_n}.
$$
This gives the anisotropy due to natural width
$$\biggl(\frac{\Delta T}{T}\biggr)_{n}=\frac{\alpha}{6}\sqrt{k-B_2}
\frac{n}{n+\frac{1}{2}}
                                                                          \eqno(33)
$$
where $\alpha$ is the fine structure constant, which at the GUT
scale is estimated within the range \cite{K}
$$
\alpha\sim{\frac{1}{25}\div \frac{1}{40}}.
                                                                     \eqno(34)
$$
The observational constraint (28) puts an upper limit on the model
parameter $k-B_2$
$$
\sqrt{k-B_2}\leq \frac{6}{\alpha} 10^{-5}.
                                                                      \eqno(35)
$$
For $\alpha$ from the range (34) this gives
$$
k-B_2 < (2.3\div 5.8)\cdot 10^{-6}.                    \eqno(36)
$$
The qualitative estimates (31), (36) allows us to conclude that
the observational constraint (28) restricts the value of the model
parameter $k-B_2$ in rather narrow range around $10^{-6}$, which
leads to some preliminary predictions concerning the quantum birth
of a hot universe.

For some value of $k-B_2$ from the admissible range, say,
$k-B_2\simeq{3\cdot 10^{-6}}$,  three cases are possible:

i) A closed universe, $k=1$, $B_2\simeq {(1-3\cdot 10^{-6})}$,
born in the presence of strings with a positive deficit angle (or
other quintessence with the equation of state $p=-\varepsilon/3$)
whose density is comparable to the vacuum density
$\varepsilon_{vac}$.

ii) An open universe, $k=-1$, $B_2\simeq {-(1+3 \cdot 10^{-6})}$,
born due to the presence of strings with a negative deficit angle,
strings density  $\varepsilon_{str}$ is comparable to
$\varepsilon_{vac}$.

 iii) Most plausible case - a flat universe, $k=0$,
$B_2\simeq{-3 \cdot 10^{-6}}$, arising from an initial
vacuum-dominated fluctuation with a small admixture of strings
with a negative deficit angle,
$\varepsilon_{str}<<\varepsilon_{vac}$.

The Friedmann equations governing the classical evolution of a
universe after tunnelling, read
$$
\dot{a}^2=\frac{8\pi G a^2}{3c^2}\left (\varepsilon_{vac} +
\varepsilon_{\gamma} \right) -\left(k-B_2\right)c^2,
$$
$$
\ddot{a}=-\frac{4\pi
G}{3c^2}\left(-2\varepsilon_{vac}+\varepsilon_{\gamma}
+3p_{\gamma}\right)                                   \eqno(37)
$$
where the dot denotes differentiation with respect to the
synchronous time.

The general constraint $E_n<U_{max}$ restricts
 the radiation
 energy density $\varepsilon_{\gamma}$ by
$$
\frac{\varepsilon_{\gamma}}{\varepsilon_{vac}} < \frac{1}{16}
                                                             \eqno(38)
$$
so that the situation at the beginning of the classical evolution
is plausible for inflation: strings (as any matter with the
equation of state $p=-\varepsilon/3$) do not contribute to the
acceleration, while the de Sitter vacuum ($p=-\varepsilon_{vac}$)
provides a huge initial expansion.

For the vacuum of GUT scale $E_{GUT}\sim {10^{15}}$ GeV the
temperature at the beginning of classical evolution is estimated
within the range
$$
0.4\cdot 10^{13}\rm GeV\leq \Theta\leq 0.3\cdot 10^{15}\rm GeV.
                                                                     \eqno(39)
$$
The lower limit on the temperature is close to the values
predicted by reheating theories \cite{reheating}. The upper limit
is far from the monopole rest energy, so that the problem of the
monopole abundance does not seem to not appear in this model.

\vskip0.1in

Now let us estimate the probability of quantum birth of a hot
universe with the parameters restricted by (28). The penetration
factor is given by the Gamow formula
$$
D=\exp\left(-\frac{2}{\hbar}|\int\limits_{a_1}^{a_2}\sqrt{2m_{pl}
(E-U)}\,da|\right).
                                                                   \eqno(40)
$$
For the potential (10) in the range of $n$ satisfying (25), this
gives
$$
D=\exp\left\{-\frac{2}{3} \left(\frac{a_0}{l_{Pl}}\right)^2
(k-B_2)^{3/2} +(2n+1)+I\right\}
                                                                        \eqno(41)
$$
where $I<10^{-2}(2n+1)$.

Near the maximum of the potential (10) the penetration factor is
calculated  using the approximation (16) which gives
$$
D_1=\exp\left\{-\frac{\pi}{4\sqrt{2(k-B_2}}
\left|\frac{(k-B_2)^2}{4}-B_4\right|
\left(\frac{a_0}{l_{pl}}\right)^2\right\}.
                                                                        \eqno(42)
$$

Comparing the penetration factors (41) and (42) with taking into
account restrictions on $k-B_2$ and $B_4$, we see that more
probable is the quantum birth of a universe from the levels with
quantum numbers $n$ from the range (25) corresponding to the
harmonic oscillator wing of the potential (10).

Formulae (41)-(42) evidently satisfy the WKB approximation since
$(a_0/l_{Pl})^2\sim{10^{16}}$ for the GUT scale
$E_{GUT}\sim{10^{15}}$ GeV, while the model parameter $k-B_2$ is
restricted by (31) and (36). For the values of this parameter
compatible with observational constraint (28), the probability of
tunnelling is estimated as
$$
D_{from ~a ~level}\sim \exp{\left(-\frac{2}{3}\cdot 10^7\right)},
\eqno(43)
$$
while the probability of the quantum birth of a universe from
nothing is estimated for the same scale $E_{GUT}$ as
\cite{V,F5,us2001}
$$
D_{from  ~nothing}\sim \exp{\left(-\frac{2}{3}\cdot
10^{16}\right)}.                   \eqno(44)
$$

\vskip0.1in

\section{Conclusions}
The main conclusion is the existence of the
lower limit on the temperature of a universe born in a tunnelling
event. A quantum fluctuation giving rise to a quantum universe
cannot in principle have a zero temperature, because its zero-level
energy has a non-zero value given by Eq.(19) for $n=0$, which is a
zero-point vacuum mode $\hbar \omega /2$. Minimal zero-level
energy puts a lower limit on a temperature of a universe arising
as a result of quantum tunnelling,  which is close to the values
predicted by reheating theories.

The upper limit on the temperature for the GUT scale vacuum
is far from the monopole rest mass, so
the problem of monopole abundance does not arise in this model.

The probability of a quantum birth from a level of non-zero energy is much bigger than
the probability of a quantum birth from nothing at the same energy scale.

The model predicts the quantum birth of the GUT-scale hot universe
with the temperature consistent with reheating theories, and
temperature fluctuations compatible with the observed CMB
anisotropy. The model does not predict the monopole abundance for
the universe born from a level of quantized temperature. Quantum
cosmology proves thus to be able to make proper predictions
concerning direct observational consequences.

\section{Acknowlegement}
This work was supported by the Polish Committee for Scientific
Research through the grant 5P03D.007.20 and through the grant for
UWM.


\end{document}